\documentclass[aps,prd,eqsecnum,nofootinbib,twocolumn,floatfix]{revtex4}
\usepackage{graphicx}
\usepackage{dcolumn}
\usepackage{times,mathptm}
\usepackage{subfigure}
\newcommand{\beq}{\begin{equation}}
\newcommand{\eeq}{\end{equation}}
\newcommand{\bea}{\begin{eqnarray}}
\newcommand{\eea}{\end{eqnarray}}

\renewcommand{\d}{\delta}

\renewcommand{\L}{\Lambda}
\renewcommand{\b}{\beta}
\renewcommand{\a}{\alpha}

\renewcommand{\k}{\kappa}

\newcommand{\g}{\gamma}
\newcommand{\n}{\nu}
\newcommand{\m}{\mu}
\renewcommand{\r}{\rho}
\newcommand{\bx}{{\mathbf{x}}}
\newcommand{\by}{{\mathbf{y}}}

\newcommand{\s}{\sigma}

\newcommand{\tQ}{\widetilde{Q}}

\newcommand{\qb}{\overline{q}}

\newcommand{\oh}{{\textstyle{\frac{1}{2}}}}

\newcommand{\dg}{\dagger}
\newcommand{\non}{\nonumber}

\newcommand{\rf}[1]{(\ref{#1})}
\newcommand{\ra}{\rightarrow}
\newcommand{\pa}{\partial}

\newcommand{\FIGURE}[2][v]{\begin{figure}[#1]#2
              \end{figure}}

\begin{document}

\title{On the Ambiguity of Spontaneously Broken Gauge Symmetry}

\author{W. Caudy}
\author{J. Greensite}
\affiliation{Physics and Astronomy Dept., San Francisco State
University, San Francisco, CA~94132, USA}
\date{\today}
\begin{abstract}
Local gauge symmetries cannot break spontaneously, according
to Elitzur's theorem, but this leaves open the possibility of breaking some 
global subgroup of the local gauge symmetry, which is typically the
gauge symmetry remaining after certain (e.g.\ Coulomb or Landau)
gauge choices.  We show that in an SU(2) gauge-Higgs system such 
symmetries do indeed break spontaneously, but the location of the 
breaking in the phase diagram depends  
on the choice of global subgroup.    The implication is that there is no
unique broken gauge symmetry, but rather many symmetries which break 
in different places. The problem is to decide which, if any, of these gauge 
symmetry breakings is associated with a transition between physically different,
confining and non-confining phases.  Several proposals $-$ Kugo-Ojima, 
Coulomb, and monopole condensate $-$ are discussed.
\end{abstract}

\maketitle
%
% Section <I
%
\section{Introduction}\label{Intro}

      Most introductory treatments of the Higgs mechanism teach that the 
spontaneous breaking of a gauge symmetry is signaled by the non-vanishing expectation 
value of a Higgs field.   Such introductory discussions occasionally
overlook the fact that local gauge symmetries \emph{cannot} break 
spontaneously, according to a celebrated theorem by Elitzur \cite{Elitzur}, and in the absence
of gauge-fixing the VEV of the Higgs field $\phi $ is rigorously zero, no matter what the form
of the Higgs potential.  In contrast, in a unitary gauge which fixes the gauge symmetry
completely, it can happen instead  that
$\langle \phi \rangle \ne 0$, again irrespective of the Higgs potential.   The point is that
only a global subgroup of a local gauge symmetry can break spontaneously, and the
order parameter for this symmetry breaking must transform non-trivially under the
global subgroup, but remain invariant under arbitrary local gauge transformations.  Local gauge
transformations, of course, can vary independently at each spacetime point; this is the
feature which is crucial to the Elitzur theorem, and the number of parameters specifying a local
gauge transformation grows with spacetime volume.  By a ``global" subgroup we mean only 
that gauge transformations in the subgroup depend on a finite and fixed number of parameters 
which is independent of volume;  such global transformations are not necessarily constant in spacetime.

     One way of constructing appropriate order parameters is via a gauge choice, which 
leaves the desired global symmetry unfixed.  Coulomb and Landau gauges are 
examples of such gauge choices.  Since the local but not the global gauge freedom has 
been gauged away, the Higgs field (and other local observables) can serve as 
order parameters for breaking of the remaining gauge symmetry.  An alternative approach is to build the 
gauge choice into the definition of the order parameter, rendering it invariant under 
local, but not global, transformations.  For example, instead of computing the VEV of 
the Higgs field in, e.g., Coulomb gauge, one could compute the VEV of the non-local operator
\beq
           \Phi(x;A) = g(x;A) \phi(x)
\eeq
where $g(x;A)$ is a field-dependent gauge 
transformation which takes the given $A$-field into Coulomb gauge.  In an abelian 
theory with an infinite spatial volume this transformation can be derived explicitly, 
and the result is
\beq
          \Phi(\bx,t;A) =  \exp\left[i\int d^3y ~ A_k(\by,t) \pa_k 
                  {1\over 4\pi |\bx- \by|}\right] \phi(\bx,t)
\label{e-charge}
\eeq
This is the Dirac construction \cite{Dirac}.  The operator $\Phi$ is invariant
under local gauge transformations which go to the identity at spatial infinity,
but transforms as a charged operator under spatially constant gauge transformations.
It is easy to see that the VEV of $\phi$ in Coulomb gauge is the same as the 
VEV of $\Phi$ evaluated without gauge fixing.  A similar construction can be 
made in the Landau gauge. 

     It is important to recognize that different gauge choices, and even different
order parameters in the same gauge, single out different global subgroups of
the full gauge symmetry.  In Landau gauge there is of course a remnant gauge
symmetry under spacetime-constant gauge transformations $g(x)=g$.  There is also,
in this gauge, an invariance with respect to certain spacetime-dependent transformations.
For the SU(2) group, with 
\beq
   g(x) = \exp[i\L^a(\epsilon;x) \oh \s_a] 
\eeq
and $\L(\epsilon;x)$ linear in the infinitesmal parameters $\epsilon_\m^a$, these
spacetime-dependent transformations can be worked out in a power series expansion in the 
coupling ${\mathbf g}$ \cite{Hata}.  To first order, $\L$ is given by
\beq 
      \L^a(\epsilon;x) =   \epsilon_{\m}^a  x^{\m}  
              - {\mathbf g}{1\over \pa^2}  (A_\m \times \epsilon_\m)^a + O({\mathbf g}^2)
\label{ghata}
\eeq 
If the Higgs field has an expectation value in Landau gauge, both the spacetime constant and
spacetime dependent global symmetries are broken.    The spacetime-dependent global symmetry \rf{ghata}
singled out in Landau gauge is not a global symmetry in Coulomb gauge (although a different
but analogous symmetry could be constructed).  Symmetry with respect to the spacetime-constant
transformations $g(x)=g$ is a remnant symmetry in both Landau and Coulomb gauges, but in
Coulomb gauge there is a much larger invariance 
with respect to transformations which are constant in space, but not in time, i.e.\ $g(\bx,t)=g(t)$. 
Suppose we single out two specific times, e.g.\ $t=0$ and $t=T$.  The trace Tr$[L]$ of
a timelike Wilson line
\beq
           L(\bx,T) = P\exp\left[i\int_{0}^{T} dt A_0(\bx,t)\right]
\label{L}
\eeq
is invariant under gauge transformations which are constant in space and time, 
\beq
              \mbox{Tr}[L(\bx,T)] =  \mbox{Tr}[g L(\bx,T) g^\dg]
\eeq 
and is therefore insensitive to the spontaneous breaking of that symmetry.  But this observable
is not invariant under the group of transformations which are constant in space, but
independent at times $t=0$ and $t=T$
\beq
              \mbox{Tr}[L(\bx,T)] \ne  \mbox{Tr}[g(0) L(\bx,T) g^\dg(T)]
\eeq 
This means that  $\langle \mbox{Tr}[L] \rangle$ in Coulomb gauge probes the breaking of a 
global gauge symmetry which is different from the symmetry probed by $\langle \phi \rangle$ in 
Landau gauge.\footnote{The Coulomb gauge remnant symmetry $g(t)$ is local in time, and 
if we consider timelike Wilson lines running from $t=t_0$ to $t=t_0+T$, then the Elitzur theorem 
guarantees that $\mbox{Tr}[L]$ would vanish if averaged over all $t_0$, as well as all 3-space 
positions $\bx$.   What happens is that is that Wilson lines can have a non-vanishing average
at fixed $t_0$, because on a timeslice the symmetry is global and can break spontaneously, 
but these spatial averages are in general different at different $t_0$, and must cancel upon averaging
over $t_0$.} 

The question which naturally arises is whether the spontaneous breaking of different global subgroups
of the local gauge symmetry, associated with different gauge choices and/or order parameters, 
occur at the same location in the space of coupling constants.  If not, then there is a certain ambiguity
in the notion of gauge symmetry breaking; precision requires specifying the particular global subgroup 
which is actually broken.   

      Assuming that different subgroups break in different places, the next question
is which (if any) of the various global subgroups is associated, upon symmetry breaking, with a transition 
to a physically different phase.   In particular, the breaking or restoration of which subgroup is associated with 
the transition from a confinement phase to some non-confining phase?   As it happens, a number of different
of approaches to the confinement problem, discussed below, associate confinement with the symmetric
(or broken) realization of different global gauge symmetries.   If these symmetries break in different places, 
it raises the obvious question of which global gauge symmetry is the ``correct" way to characterize confinement, 
particularly when global center symmetry (which is \emph{not} a gauge symmetry) is broken by matter fields
      
        In this article we will investigate the possible ambiguity of
gauge symmetry breaking in the context of a gauge-Higgs theory on the lattice, 
with a fixed-modulus Higgs field in the fundamental color representation.  For the 
SU(2) gauge group, the Lagrangian can be written in the form \cite{Lang}
\begin{equation}
     S = \b \sum_{plaq} \oh \mbox{Tr}[UUU^\dg U^\dg] \non \\
       + \gamma \sum_{x,\m} \oh \mbox{Tr}[\phi^\dg(x) U_\m(x)
\phi(x+\widehat{\m})]
\label{ghiggs}
\end{equation}
with $\phi$ an SU(2) group-valued field.  Investigations \cite{FS,OS} of this model, carried out
many years ago, revealed an important and surprising feature:
Consider two points $(\b_1,\g_1)$ and $(\b_2,\g_2)$ in the $\b-\g$ phase diagram,
with $(\b_1,\g_1) \ll 1$ deep in the ``confinement" (strong-coupling) regime, and $(\b_2,\g_2) \gg 1$ deep
in the Higgs regime.   Then according to a
result due to Fradkin and Shenker \cite{FS} (which was based on an earlier theorem of 
Osterwalder and Seiler \cite{OS}), there is a path in the phase diagram connecting the
two points, such that the expectation value of any local gauge-invariant observable, or product
of such observables, varies analytically along the path.  This means that there is no 
thermodynamic phase transition which entirely isolates the Higgs phase from a confinement phase. 
Subsequent numerical work \cite{Lang,ghiggs} ruled out a massless phase, and indicated
the phase structure sketched in Fig.\  \ref{fradshenk}, with a line of first order transitions
(or possibly just a line of rapid crossover) which ends at around $\b=2,\g=1$, consistent with
the Fradkin-Shenker-Osterwalder-Seiler theorem.   Above the transition line, at large $\b$,
the dynamics is clearly that of a Higgs phase, with a massive spectrum, no linear Regge
trajectories, no flux tube formation, and only Yukawa-type potentials between static color 
charges.   On the other hand, at small values of $\gamma$, the theory is reminiscent of 
real QCD with dynamical fermions.    In this coupling regime we have flux tube formation and a linear
potential over some finite distance range, followed by string breaking via scalar particle
production.

\FIGURE[t]{
\centerline{\rotatebox{270}{\includegraphics[width=6truecm]{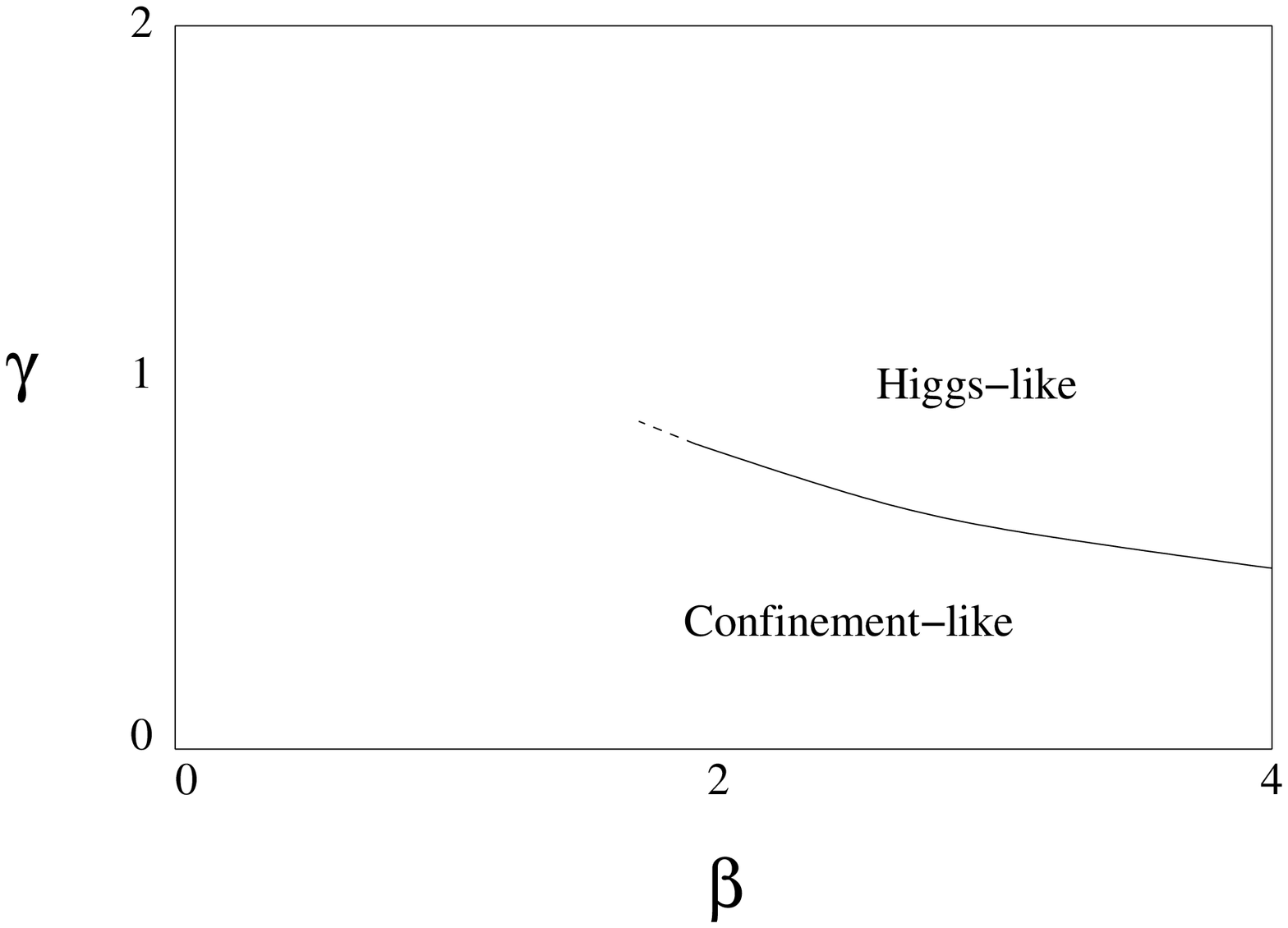}}}
\caption{Schematic phase diagram of the SU(2) gauge-Higgs system.  
The solid line is a line of weak first-order phase
transitions.} 
\label{fradshenk}
}

    One of the things that we learn from the Fradkin-Shenker work is that the Higgs phase cannot
be distinguished from the confinement phase by so-called ``color confinement" in the
asymptotic particle spectrum.  It is always possible to choose a path, from the confinement to the Higgs 
regime, such that all local gauge-invariant observables, products of such observables, and (in particular)
the free energy, vary analytically
along the path, and this behavior is incompatible with an abrupt, qualitative change in the spectrum.
Asymptotic particle states are therefore color singlets throughout the phase diagram.  
In the absence of a massless, Coulombic regime, color is \emph{always} 
screened by the fundamental-representation Higgs field, whether this screening is
viewed as a string-breaking effect, or as the rearrangement of a condensate in the 
neighborhood of a color charge.

     We then return to the basic question:  {\it In the absence of a thermodynamic separation,
can the spontaneous breaking of a gauge symmetry distinguish unambiguously between 
the Higgs and confinement phases?}   To address this question,  we will map out the
location of the breaking of remnant global gauge symmetries in the Coulomb and Landau
gauges. It will be found that these transitions coincide, within the accuracy of our data,
along the thermodynamic transition line at $\b > 2$.  But away from that line, at $\b<2$, the 
transitions are found to diverge from one another.   This result ties in with an earlier work \cite{GO1}
in the gauge-Higgs theory, comparing the line of gauge symmetry breaking in Coulomb gauge 
with the line of center vortex percolation/depercolation (a ``Kert{\'e}sz" line \cite{Kertesz}), which were thought to be
identical \cite{GOZ}.  In fact, the Coulomb gauge and percolation transition lines also coincide with
the thermodynamic transition line at $\b>2$, but diverge from one another at lower $\b$. 
Percolation transitions at finite temperature, for other types of topological objects in electroweak gauge theory
and QCD, were also discussed in ref.\ \cite{Maxim}, where it was pointed out (in the second article cited) 
that the precise location of the Kert{\'e}sz line depends on the type of object studied.  Of course, 
spontaneous breaking of a gauge symmetry and a percolation transition are in principle very different things, and
the result in ref.\ \cite{GO1} leaves open the question of whether or not spontaneous breaking of 
different global gauge symmetries coincide.   

     In the next section we will discuss the order parameters for confinement in three
different approaches:  (i) the Kugo-Ojima criterion (covariant gauges); (ii) Coulomb confinement
(Coulomb gauge); and (iii) dual superconductivity.  Each of these
order parameters is sensitive to the breaking of a different global gauge symmetry.  In section 3
we present our data for global gauge symmetry breaking, in Landau and Coulomb gauges, in the
SU(2) gauge-Higgs model.  Symmetry breaking associated with the third order parameter, which
is less straightforward to implement numerically, will be
reserved for a later study.  Section 4 contains discussion and conclusions.

\section{Order Parameters for Confinement}

        In gauge theories with a non-trivial center symmetry, there is no difficulty
in distinguishing qualitatively between the confinement phase and the Higgs phase,
or between confinement and a high temperature deconfined phase.  The
vanishing of Polyakov lines, the large-volume behavior of the vortex free energy, 
and the non-vanishing of string tensions extracted from 
fundamental representation Wilson loops, all serve as
appropriate, consistent, and gauge-invariant signals of the confinement phase \cite{review}.
A transition away from the confinement phase is always accompanied by the
spontaneous breaking of the global center symmetry, and non-analytic behavior
in the free energy.
But the situation is much less clear when there are dynamical matter fields
in the fundamental representation of the gauge group, as in real QCD.
When global center symmetry is broken explicitly, Polyakov lines are
non-zero, and Wilson loops fall off asymptotically with a perimeter-law behavior,
as in a Higgs phase.   The question is whether there is some other symmetry
which can distinguish the confinement phase from other massive phases.  We will
discuss three proposals, each of which could  potentially identify the confined phase even
in the absence of a global center symmetry in the Lagrangian.

\subsection{The Kugo-Ojima criterion}

     Kugo and Ojima \cite{Kugo1} begin with an equation satisfied by the 
conserved color current $J_\m^a$ in covariant gauges
\beq
           J_\m^a = \pa^\n F_{\m \n}^a + \{Q_B,D_\m^{ab} \overline{c}^b\}
\eeq
where $c,\overline{c}$ are the ghost-antighost fields with $Q_B$ the BRST charge, and
also introduce the function $u^{ab}(p^2)$, defined by the expression
\bea
 \lefteqn{u^{ab}(p^2) \left( g_{\m\n} - {p_\m p_\n \over p^2} \right) = }
 \non \\
           & &   \int d^4 x ~ e^{ip(x-y)} \langle 0|T[D_\m c^a(x) 
           g(A_\n \times \overline{c})^b(y) | 0 \rangle
\eea
They then show that the expectation value of color charge in any physical
state vanishes  
\beq
           \langle \mbox{phys}\ |Q^a |\mbox{phys}\rangle = 0
\label{Qa}
\eeq 
providing that (i) remnant symmetry with respect to spacetime-independent
gauge transformations is unbroken; and (ii) the following condition is satisfied: 
\beq
           u^{ab}(0) = -\d^{ab}
\label{kg}
\eeq
This latter condition is the Kugo-Ojima confinement criterion, and it implies that the ghost propagator
is more singular, and the gluon propagator less singular, than a simple pole at $p^2=0$ \cite{Kugo2}.  
A number of efforts have focussed on verifying this condition (or its corollaries) both
analytically \cite{Alkofer} and numerically \cite{Nakajima}.   

      It turns out that the Kugo-Ojima condition \rf{kg} is itself tied to the unbroken
realization of remnant gauge symmetry in covariant gauges (such as Landau gauge).  
We have already noted that in Landau gauge there is a remnant group of spacetime-dependent gauge 
transformations, given in eq.\ \rf{ghata},  which preserves the Landau gauge condition.  It was shown by 
Hata in ref.\ \cite{Hata} (see also Kugo in ref.\ \cite{Kugo2}) that the condition \rf{kg} is a necessary 
(and probably sufficient) condition for the unbroken realization of the residual spacetime-dependent 
symmetry \rf{ghata}, while an unbroken, spacetime independent
symmetry is required, in \emph{addition} to \rf{kg}, for the vanishing of 
$\langle \psi |Q^a |\psi \rangle$ in physical states. 

   Thus the Kugo-Ojima scenario requires the 
full remnant gauge symmetry in Landau gauge, i.e.\ both the spacetime dependent and the spacetime 
independent  residual 
gauge symmetries must be unbroken.  Both of these symmetries 
are necessarily broken if a Higgs field aquires a VEV in Landau gauge.

\subsection{The Coulomb gauge criterion}

      The criterion for confinement as the unbroken realization of remnant gauge symmetry
in Coulomb gauge was first put forward by Marinari et al.\ in ref.\ \cite{Marinari}; the idea
was elaborated and studied numerically in ref.\ \cite{GOZ}.  The criterion can be
motivated as follows:  In Coulomb gauge it is simple to construct color non-singlet 
physical states; an example is
\beq
             \Psi_q^a = q^a(x) \Psi_0
\eeq
where $\Psi_0$ is the vacuum state in Coulomb gauge, and $q^a(x)$ is a heavy
quark operator.  Whereas the aim of the Kugo-Ojima approach is to prove that the
space of physical states consists of only color singlets, the goal in Coulomb gauge is
to prove that color non-singlet states have an energy which is infinite above the
vacuum.   For heavy quarks, with a lattice regularization understood, we define
\bea
            G(T)  &=& \langle \Psi^a_q | e^{-(H-E_0)T}|\Psi^a_q\rangle
\non \\
                 &\propto&  \left\langle \mbox{Tr}\bigl[L(\bx,T)\bigr]  \right\rangle                   
\label{sum}
\eea
The energy of the charged state $\Psi_q$ is infinite if $G(T)=0$, i.e.\
$\langle \mbox{Tr}[L] \rangle =0$, and finite otherwise.  This means that
the Coulombic field energy of an isolated charge is infinite if the
remnant global gauge symmetry associated with the pair of spatially homogeneous 
transformations $g(0),~g(T)$ is unbroken.  Conversely, an isolated
color charge has finite energy if this remnant symmetry is spontaneously
broken.

  One can also show that the instantaneous color Coulomb potential between
quark-antiquark color charges is given
by the logarithmic derivative of the correlator of timelike lines \cite{GO}
\beq
          V_{coul}(R) = - \lim_{T\ra 0} {d\over dT} \log\Bigl[\mbox{Tr}
                 [L(\bx,T)  L^\dg(\by,T)] \Bigr]
\label{Vcoul}
\eeq
($R=|\bx-\by|$), and this potential is an upper bound on the static quark potential \cite{Dan}.
If $\langle \mbox{Tr}[L] \rangle \ne 0$, then $V_{coul}(R)$ is $R$-independent as 
$R\ra \infty$, and therefore non-confining.  This is a further motivation for the use of 
timelike Wilson lines, in Coulomb gauge, as an order parameter for confinement.
 
   In principle, the color Coulomb potential can reveal the confining nature of 
the vacuum even in the presence of dynamical matter fields, because of its instantaneous nature.
The color Coulomb potential derives from the non-local term in the Coulomb
gauge Hamiltonian.  When the VEV of this term is evaluated in  a state
such as
\beq
           \Psi_{q\qb} = \qb^a(\bx) q^a(\by) \Psi_0
\eeq
containing isolated quark-antiquark charges, it accounts for the energy of the associated Coulomb field
before the quark-antiquark system has evolved in time, and screening effects due to matter and/or
transverse gluon fields have set in.   This means that Tr$[L]$ can work as an order parameter
for confinement even when, as in real QCD, there exist dynamical matter fields which break
the global center symmetry.    Confinement, in this approach, is identified with the phase in which the
energy of the Coulomb field due to isolated color-charge sources diverges as the charge separation
is taken to infinity.   That also means that confinement is tied to the unbroken realization of a specific
global subgroup of the gauge symmetry, which remains after fixing to Coulomb gauge.
   
\subsection{Dual Superconductivity}
 
         It is an old idea, due originally to 't Hooft and Mandelstam, that the Yang-Mills vacuum is
a kind of dual superconductor, in which the roles of the $E$ and $B$ fields are interchanged.  It is
then electric, rather than magnetic, charges which are confined, and magnetic,
rather than electric, charges which are condensed.  Magnetic monopoles can exist in gauge theories
with compact abelian gauge groups, and an order parameter for monopole condensation, breaking
the dual U(1) gauge symmetry associated with magnetic charge conservation, was introduced
in ref.\ \cite{Adriano1}.   The order parameter $\m(\bx)$ is a monopole creation
operator, which acts on states in the Schrodinger representation by inserting a monopole field
configuration $A_i^M(y)$, centered at $\by=\bx$ i.e.
\beq
              \m(\bx) |A_i \rangle = | A_i+ A_i^M \rangle 
\label{pisa-op}Ä
\eeq
Explicitly, the operator
\beq
               \m(\bx) = \exp\Bigl[i \int d^3 y ~ A_i^M(y) E_i(y) \Bigr]
\label{m-charge}
\eeq
performs the required  insertion. In a non-abelian SU(N) gauge theory, an abelian projection gauge  
must be introduced to single out an abelian $U(1)^{N-1}$ subgroup, and $\m$ is defined in terms of the
gauge fields associated with that subgroup.  Details concerning this construction on the
lattice, and the numerical computation of  $\langle \m \rangle$, can be found in ref.\ \cite{Adriano2}.
 
    The dual U(1) gauge symmetry, in an abelian theory containing magnetic charge, is  evident
from the existence of a conserved magnetic current.  Let
\beq
            \widetilde{F}_{\m\n} = \oh \epsilon_{\m\n\a\b}F^{\a\b}
\eeq
be the dual field strength tensor.  Then
\beq
             j^M_\m = \pa^\n \widetilde{F}_{\m\n}
\eeq   
is the conserved magnetic current associated with the dual gauge symmetry.   A global 
U(1) subgroup of this local symmetry is generated by the total magnetic charge operator, 
and it is shown in ref.\ \cite{Adriano1} that the
$\m$ operator transforms like a magnetically charged object under these global symmetry
transformations.  Thus, according to ref.\ \cite{FM}, the $\m$ operator is in some sense
the dual of the Dirac construction of electrically charged operators in eq.\ \rf{e-charge}.  
If $\langle \m \rangle \ne 0$, this signals both monopole condensation, and the 
associated breaking of a global U(1) gauge symmetry in the dual gauge theory. 

As with the Kugo-Ojima and Coulomb conditions, monopole condensation  
can be put forward as a confinement criterion whether or not there are dynamical matter fields
in the theory,
and whether or not global center symmetry is broken.  Like the other two criteria, the
condition that $\langle \m \rangle \ne 0$ is tied to the spontaneous breaking of a global subgroup of
some local gauge symmetry.  Although we will not directly investigate the $\m$ operator here, 
we believe that the general issues we are going raise in connection with 
spontaneous gauge symmetry breaking 
apply to this approach as well.  The Kugo-Ojima,
Coulomb gauge, and dual-superconductor order parameters for confinement 
are all very well motivated,
and each is associated with the way in which some global gauge symmetry is realized.    
But what if, in practice,      
these criteria disagree with one another in identifying the boundary between the Higgs and confinement
phases?  Which symmetry is the ``right" one, in terms of identifying physically distinct phases?
This question is reserved for the concluding section; we first need to show that the location of
gauge symmetry breaking is, in fact, dependent on the choice of the global subgroup.  

\begin{figure*}[tbh]
\begin{center}
\subfigure[] % caption for subfigure a
{
    \label{2p2e}
    \includegraphics[width=8truecm]{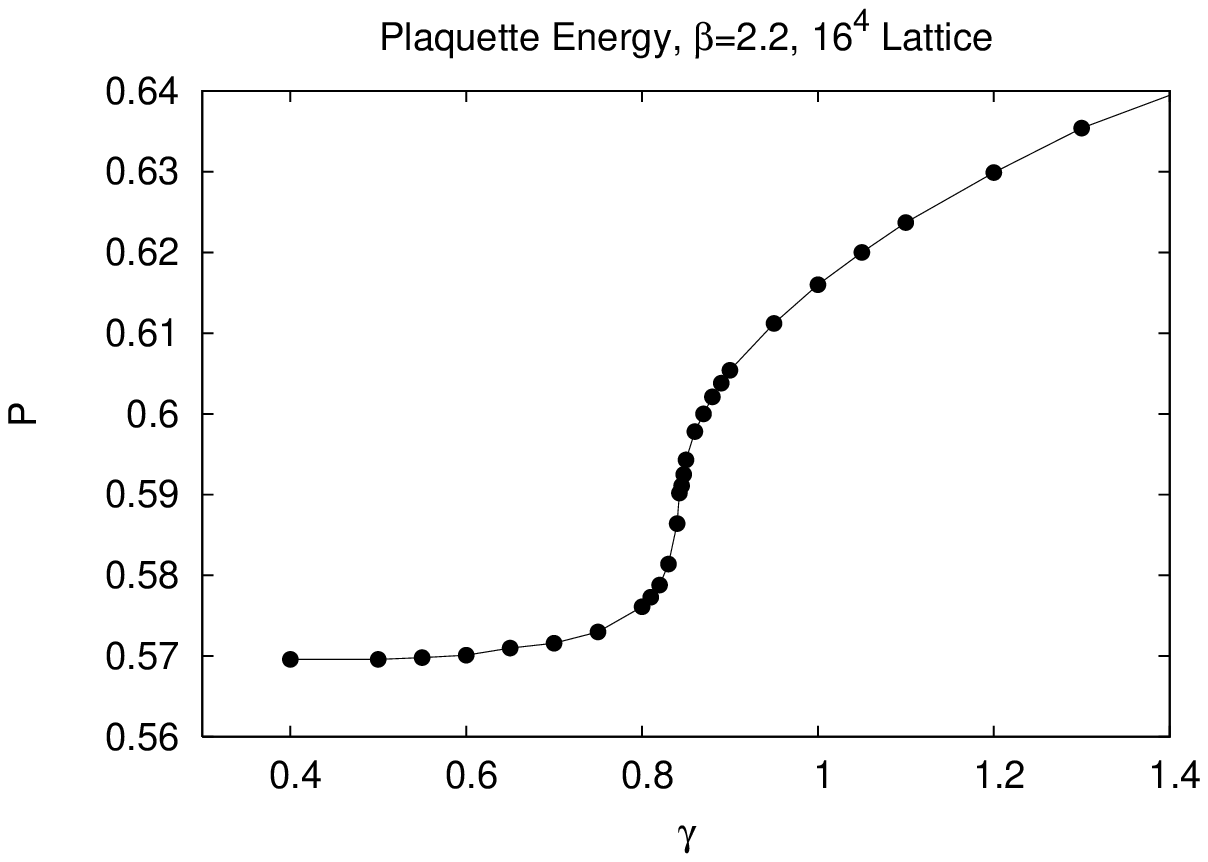}
}
\hspace{0.25cm}
\subfigure[] % caption for subfigure b
{
    \label{1p2e}
    \includegraphics[width=8truecm]{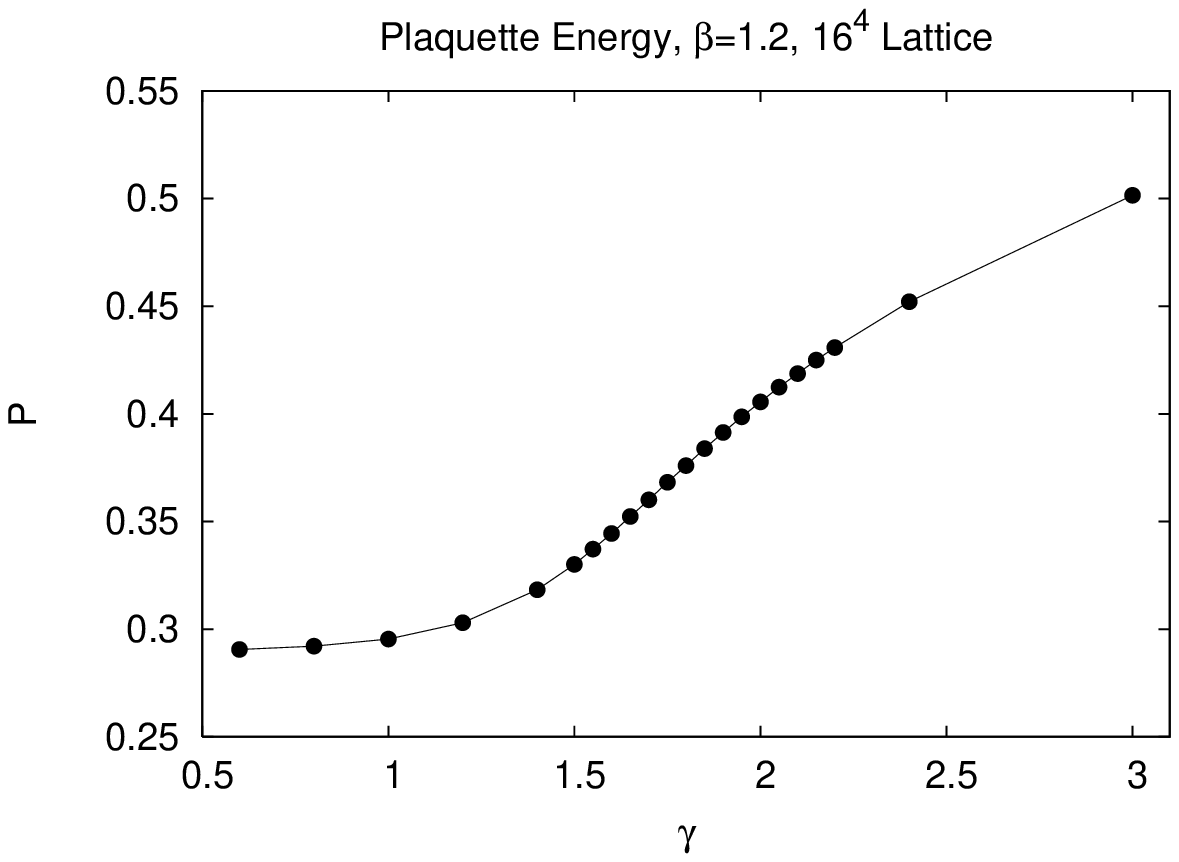}
}
\end{center}
\caption{Plaquette expectation value $P$ vs.\ Higgs coupling $\g$ at gauge couplings:
(a) $\b=2.2$, either a sharp crossover or a weak first-order transition is seen 
at $\g=0.84$, and (b) $\b=1.2$, no transition is evident. The figures are taken
from ref.\ \cite{GO1}. }
\label{fig0} % caption for the whole figure
\end{figure*}

\section{Remnant symmetry breaking in Coulomb and Landau gauges}

    The order parameter for remnant symmetry breaking in 
Landau gauge is straightforward.  In Landau gauge, the remnant symmetry is broken if
the magnitude of the spatial average of the Higgs field is non-zero in the infinite volume limit.
Denoting the spatial average as
\beq
       \widetilde{\phi} = {1\over V} \sum_x \phi(x)
\eeq
we define\footnote{This operator was applied previously by Langfeld \cite{Kurt}
to determine global gauge symmetry breaking transitions in SU(2) and SU(3) gauge-Higgs theories,
fixed to Landau gauge.  The models studied in that work used Higgs fields of variable modulus, 
so the transition points are not directly comparable to our data.} 
\bea
               \tQ_L &=& \oh \mbox{Tr}[\widetilde{\phi} \widetilde{\phi}^\dg]
\non \\
                 Q_L &=& \langle \tQ_L \rangle
\eea
where $V$ is the lattice 4-volume.  The global remnant symmetry is
unbroken iff $Q_L \ra 0$ as $V \ra \infty$.  In fact, it is easy to see that if the symmetry
is unbroken, and the Higgs field has a finite correlation length in Landau gauge, then
\beq
           Q_L \propto {1 \over V}
\eeq
whereas $Q_L \ra \mbox{const.} > 0$ as $V \ra \infty$ in the broken phase.

%%%%%%%%%%%%%%%%%%%%%%%%%
\begin{figure}[tbh]
\includegraphics[width=9truecm]{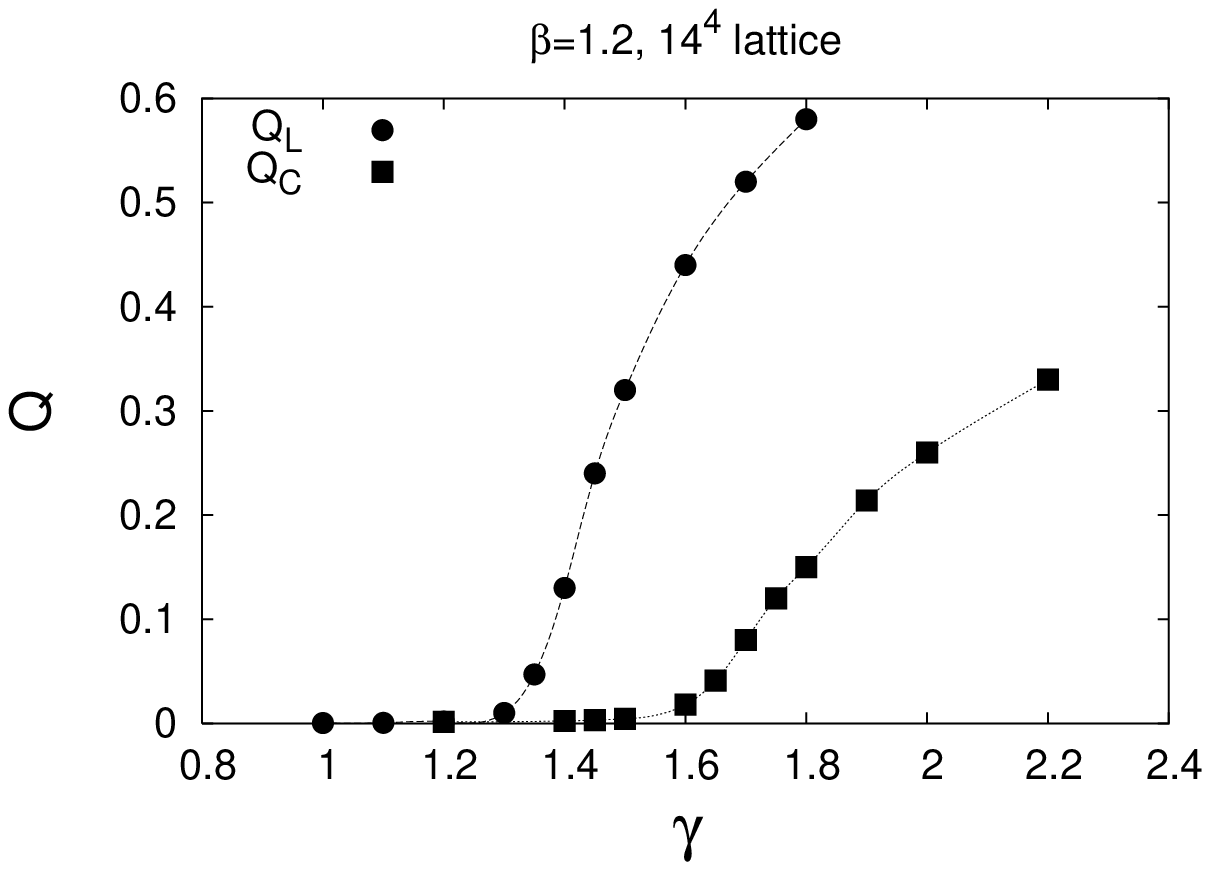}
\caption{Order parameters $Q_L$ and $Q_C$, for global gauge symmetry
breaking in Landau and Coulomb gauge respectively, vs.\ gauge-Higgs coupling
$\g$ at fixed lattice volume $14^4$ and gauge coupling $\b=1.2$.
\label{Q12} }
\end{figure}
%%%%%%%

    In Coulomb gauge there is a larger remnant gauge symmetry, in which gauge transformations
$g(\bx,t)=g(t)$ which are constant in the spatial directions can nevertheless vary in time.
We can use the timelike lattice link variables $U_0(x)$ as order parameters for this symmetry
breaking, as previously proposed in \cite{GOZ},  since Tr$[U_0]$ is sensitive to symmetry
transformations $g(t)$ which depend on $t$, but is invariant with respect to transformations
which are also constant in the time direction.  On the lattice, the logarithm of the $U_0$ correlator 
has also been used, in accordance with eq.\ \rf{Vcoul}, to calculate the color Coulomb potential 
\cite{GO}.   Denoting the spatial average of timelike links on a timeslice as
\beq
      \widetilde{U}(t) = {1\over V_3} \sum_{\bx} U_0(\bx,t)
\eeq
where $V_3$ is the 3-volume of a timeslice, we define\footnote{Note that this differs slightly
from the observable proposed in \cite{GOZ}, which defines $Q_C$ by taking the square root of the trace.}
\bea
            \tQ_C &=& {1 \over L_t} \sum_{t=1}^{L_t}  \oh \mbox{Tr}[ \widetilde{U}(t)
                               \widetilde{U}^\dg(t)] 
\non \\
             Q_C &=& \langle \tQ_C \rangle
\eea
In the unbroken phase, assuming finite-range correlations among the timelike links
at constant $t$,
\beq
           Q_C \propto {1 \over V_3}
\eeq
while $Q_C$ converges to a non-zero constant, in the broken phase, in the infinite
volume limit.
 
    The phase structure of the SU(2) gauge-Higgs model, sketched in Fig.\ \ref{fradshenk}, is
reflected in plots of the plaquette expectation value $P$ vs.\ $\g$, as shown in Figs.\
\ref{2p2e} and \ref{1p2e}, which are taken from ref.\ \cite{GO1}.  For $\b>2$, we find a sudden
rise in $P$ at some value of $\g$, as seen, e.g., in Fig.\ \ref{2p2e} for $\b=2.2$.  The data at
this coupling indicates either a weak first-order transition, at $\b=2.2,\g=0.84$, or possibly just
a sharp crossover.  The evidence for the first-order nature of the transition, for $\b$ values
above this coupling, is given in ref.\ \cite{ghiggs}.  Below $\b\approx 2$ 
(see Fig.\ \ref{1p2e} at $\b=1.2$) there is no indication,
in the $P$ vs.\ $\g$ data, of any nonanalytic behavior in the
observable, as expected from the Fradkin-Shenker-Osterwalder-Seiler theorem.

%%%%%%%%%%%%%%%%%%%%%%%%%%%
\begin{figure*}[tbh]
\begin{center}
\subfigure[] % caption for subfigure a
{
    \label{L12}
    \includegraphics[width=8.5truecm]{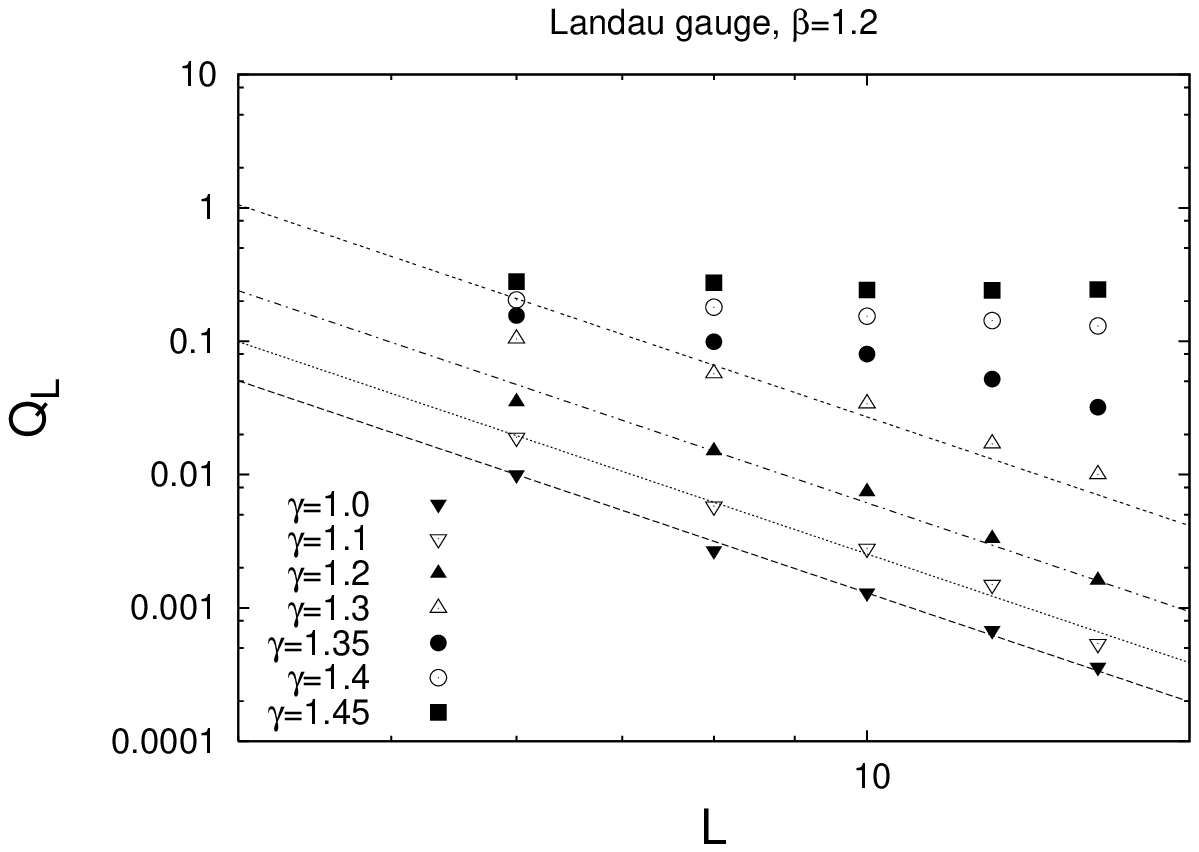}
}
\hspace{0.25cm}
\subfigure[] % caption for subfigure b
{
    \label{C12}
    \includegraphics[width=8.5truecm]{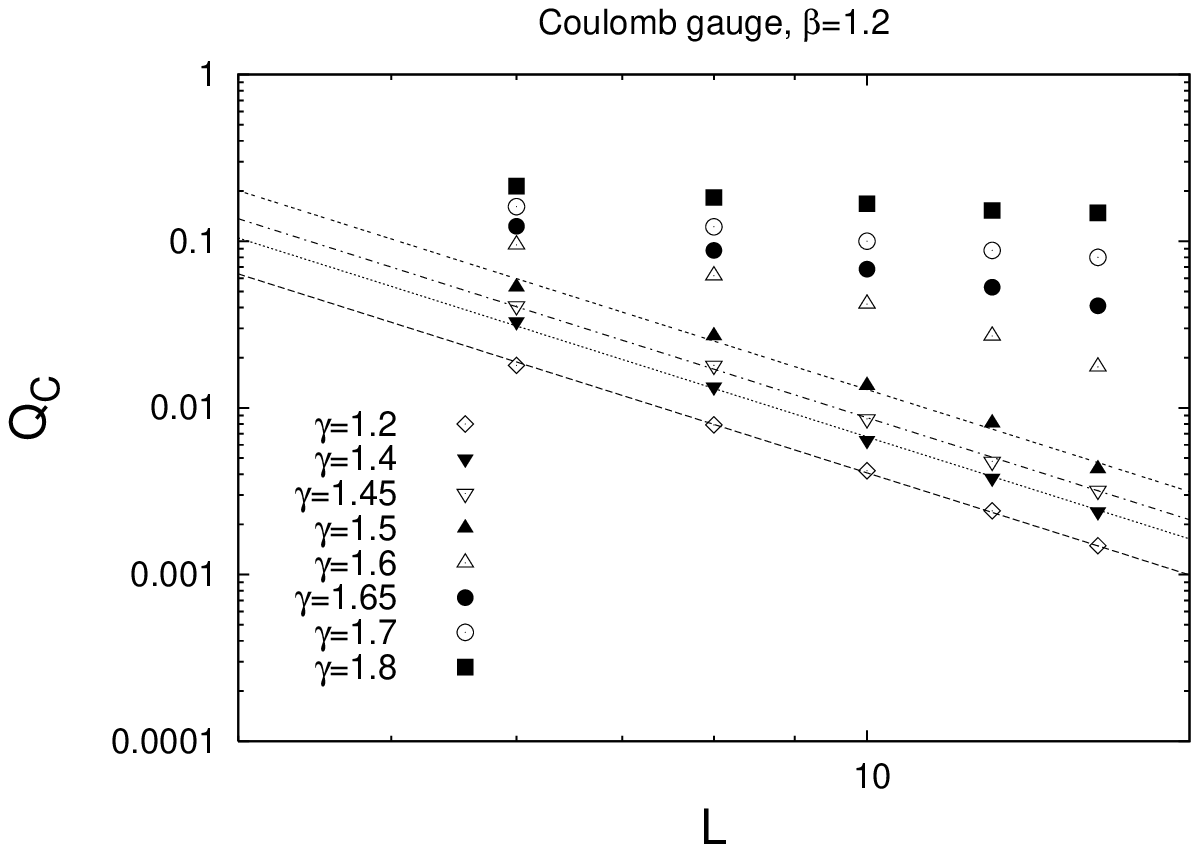}
}
\end{center}
\caption{Log-log plot of the gauge-symmetry breaking order parameters $Q_L$ and $Q_C$ vs.\
lattice extension $L$, at $\b=1.2$ and a variety of gauge-Higgs couplings $\g$,
in (a) Landau and (b) Coulomb gauges.  In the Landau and Coulomb gauges the straight lines are 
a best fit to eqs.\ \rf{QLfit} and \rf{QCfit}, respectively.} 
\label{fig1} % caption for the whole figure
\end{figure*}
%%%%%%%%%%%%%%%%%%%%%%%%%%%%%%

    We will now display our evidence that, for fixed $\b<2$, there is a transition in $Q_C$
and $Q_L$ away from zero, in the infinite volume limit, to some non-zero value, but that this
transition happens at different couplings $\g$ for the Coulomb and Landau order parameters.

    Figure \ref{Q12} is a plot of $Q_L$ and $Q_C$ vs.\ $\g$ at $\b=1.2$, on
a hypercubic lattice of volume $14^4$.    At low $\g$ both $Q_C$ and $Q_L$ are
very small, and cannot be distinguished from zero on the scale of the graph.
At some $\g$ both $Q_C$ and $Q_L$ rise rapidly away from zero, indicating
a non-zero value in the infinite volume limit.  However, this rise begins at
\emph{different} values of $\g$ for the two observables.

    Figure \ref{L12} is a log-log plot showing the dependence of $Q_L$ on the lattice extension
$L$, with $L=6,8,10,12,14$.  The coupling $\b=1.2$ is fixed, and we show results
for several $\g$ values.   The straight lines are a best fit of the data to
\beq
              Q_L = {\k \over L^4}
\label{QLfit}
\eeq
We see that the fit is quite good at the lower $\g$ values, which supports 
the extrapolation to $Q_L=0$ in the infinite volume limit.  On the other hand,
at $\g=1.45$, there is very little falloff in $Q_L$ with lattice volume, indicating
a non-zero infinite volume limit.  This means that somewhere there is a transition from a phase
of unbroken Landau gauge remnant symmetry to a broken phase.   Figure \ref{C12} shows
the same type of data for $Q_C$ at $\b=1.2$ computed in Coulomb gauge.  This time the straight lines
on the log-log plot are a best fit to
\beq
               Q_C = {\k' \over L^3}
\label{QCfit}
\eeq
Once again, the evidence supports an extrapolation to $Q_C=0$ at low $\g$, and a non-zero
value at higher $\g$, implying a transition from an unbroken to a broken phase of
Coulomb gauge remnant symmetry.  However, the actual Coulomb and Landau gauge 
transition points must be different, as we see from the fact that at $\g=1.45$ the observable
$Q_L$ is roughly $L$-independent, and therefore in the broken phase, 
while the data for $Q_C$ at this value of gamma are consistent with a $1/L^3$ falloff, 
and unbroken Coulomb gauge remnant symmetry. 

\begin{figure*}[t!]
\centering
\subfigure[] % caption for subfigure a
{
    \label{L12peak}
    \includegraphics[width=8.5truecm]{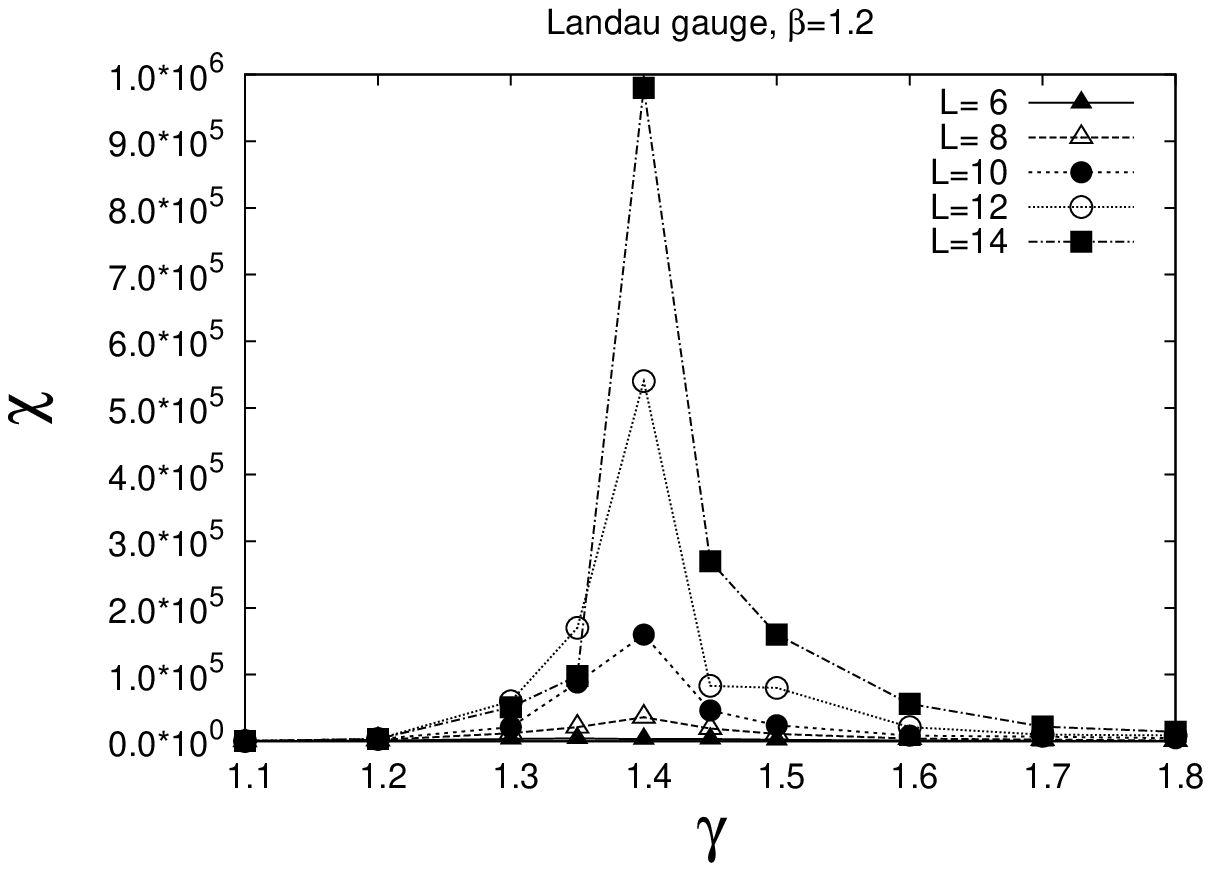}
}
\hspace{0.25cm}
\subfigure[] % caption for subfigure b
{
    \label{C12peak}
    \includegraphics[width=8.5truecm]{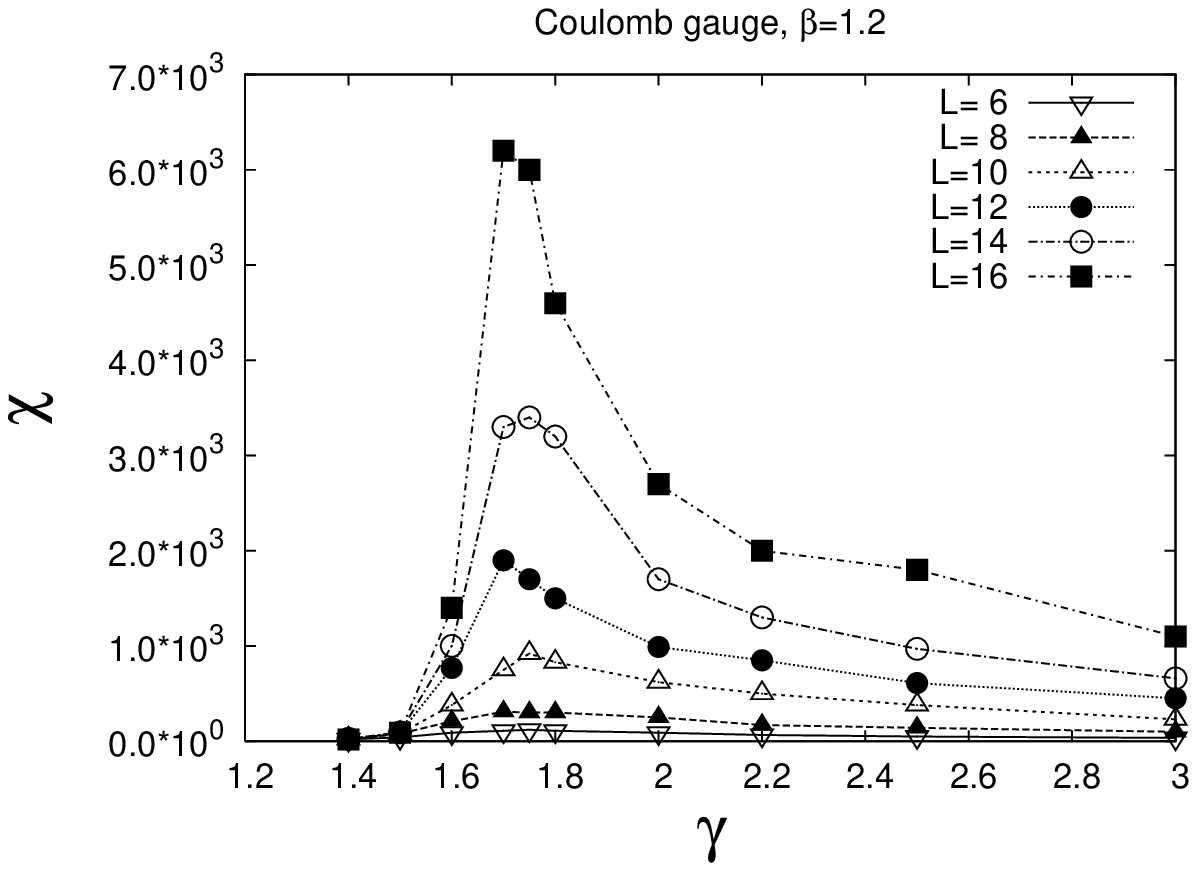}
}
\caption{Susceptibilities $\chi$ vs.\ gauge-Higgs coupling $\g$ at
fixed $\b=1.2$ and a variety of lattice volumes $L^4$.  (a) Landau gauge;
(b) Coulomb gauge.}
\label{peaks} % caption for the whole figure
\end{figure*}

    In order to improve the accuracy of our determination of the transition point, we follow
the procedure of looking for the value of $\g$ where fluctuations in the order parameter
are largest.\footnote{We will not, however, attempt a finite size scaling analysis. The
order of the transition is not especially important to us, particularly because there is,
at $\b<2$, no actual thermodynamic transition.  It is enough,
for our purposes, to establish that a transition exists, in which $Q\ra 0$ below the critical
$\g$, and $Q>0$ above the critical $\g$, in the infinite volume limit.}  
For this we define
\bea
          \chi_L &=&   V^2 \Bigl( \langle \tQ_L^2 \rangle -  Q_L^2   \Bigr)
\non \\
          \chi_C &=&   V_3^2 \Bigl( \langle \tQ_C^2 \rangle -  Q_C^2 \Bigr)
\eea
The overall volume-squared factor in these expressions is chosen to
keep $V^2 Q_L^2$ and $V_3^2 Q_C^2$ a volume-independent constant, for couplings
where $Q_{L,C} \ra 0$ in the infinite volume limit.    The results for $\chi_L$ and
$\chi_C$, respectively, at $\b=1.2$, are shown in Figs.\ \ref{L12peak} and \ref{C12peak}.
From this data we locate the remnant symmetry breaking transition points at $\g=1.4$ for 
Landau gauge, and $\g=1.7$ for Coulomb gauge, with uncertainties on the order of $0.03$.

\begin{figure}[tbh]
{\includegraphics[width=9truecm]{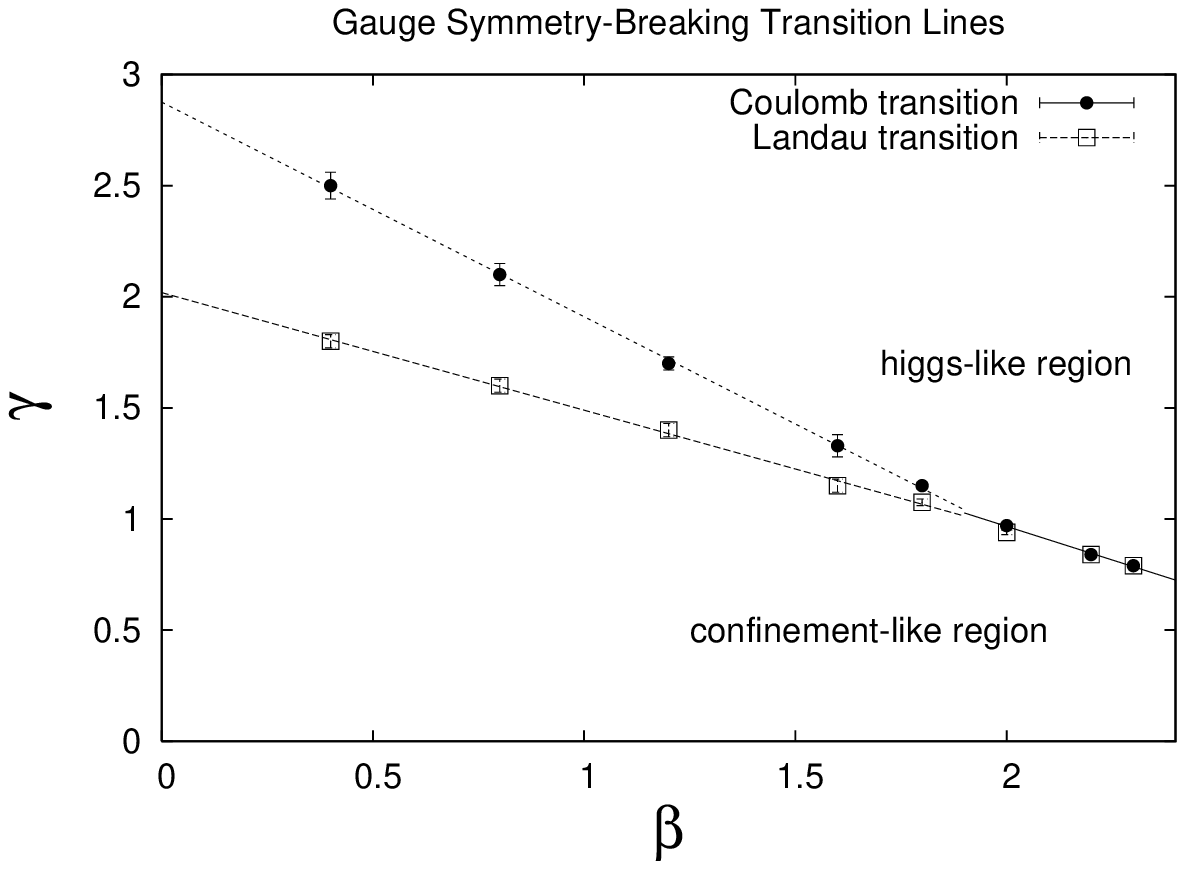}}
\caption{The location of remnant global gauge symmetry breaking in 
Landau and Coulomb gauges, in the $\b-\g$ coupling plane.} 
\label{phase}
\end{figure}

    We have applied these methods to determine the Landau and Coulomb remnant symmetry
breaking transition points at $\b=0.4,0.8,1.2,1.6,1.8,2.0,2.2,2.3$, with the results  shown in
Fig.\ \ref{phase}.    There is a clear separation of the two transition lines for $\b < 2$, where there
is no thermodynamic transition, while
at $\b > 2$ the symmetry breakings coincide with each other and with the thermodynamic
transition/crossover points, within the accuracy of our
measurements.  This is the central result of our paper.  Center vortex percolation/depercolation
transitions in the SU(2) gauge-Higgs model were investigated in ref.\ \cite{GO1}, and
it was found that at $\b \ge 2$ the percolation transition points also
coincide with the thermodynamic transitions,  while at $\b < 2$  the percolation
transitions lie above the Coulomb transition line \cite{GO1}.

\section{Discussion and Conclusions}

    We have shown that in the SU(2) gauge-Higgs model there is no
unique transition line between unbroken and spontaneously broken
gauge symmetry; instead there are different transition lines corresponding
to different global subgroups of the local symmetry.  Two subgroups
in particular, one associated with the Kugo-Ojima confinement criterion,
and the other with the confining color Coulomb potential, are found to
have distinct transitions.   The order parameters for these two symmetries
cannot both be order parameters for the transition from a
``confinement" to a Higgs phase; this seems to be a firm conclusion
of our study.   In fact, since the particle spectrum consists of only
color singlets throughout the phase diagram, and the asymptotic
string tension is zero (except at $\g=0$) throughout the phase diagram,
it is unclear in exactly what sense a transition in either of these order 
parameters is associated with a transition to or from a confined phase.  

The larger question is whether the breaking of  these or \emph{any}
global gauge symmetries necessarily indicates a transition 
between physically different phases in non-abelian gauge theory.  Of course,
gauge symmetry breaking may accompany a change of state when there
a thermodynamic phase transition.   But the question is whether 
gauge symmetry breaking
is \emph{always} accompanied by a change of physical state, even when the 
thermodynamic transition is absent.

     On the basis of the Fradkin-Shenker-Osterwalder-Seiler theorem, there
is a compelling case that no transition exists in the SU(2) gauge-Higgs
model from a Higgs phase to a physically distinct  ``confinement-like"
phase, which includes the strong-coupling region. 
If we consider any two points $(\b_1,\g_1) \ll 1$ and $(\b_2,\g_2)\gg 1$ in the
coupling constant plane, then there is always a path between them along which
the VEV of all local gauge invariant observables vary analytically, and
Green's functions constructed from such observables vary analytically.
As a consequence, the free energy and the spectrum vary analytically.
Moreover, the usual order parameters for confinement, i.e.\ the asymptotic
string tension (which vanishes) and Polyakov lines (which don't vanish) 
exhibit non-confining behavior throughout the coupling constant plane,
for any $\g > 0$.   There is simply no evidence for, and strong
evidence against,  any abrupt change separating the Higgs region from the
strong coupling confinement-like region.  So the fact that global gauge
symmetries do break spontaneously in gauge-Higgs theory at small $\b$, with different
symmetries breaking in different places in the coupling-constant
plane, makes it very unlikely that spontaneous breaking of these global gauge
symmetries necessarily correspond to a change in physical state.

      It is worth noting, in passing, that the absence of an isolated Higgs
phase in SU(2) gauge-Higgs theory also makes it clear that there is no fundamental 
distinction between string breaking by pair-production of scalar particles, and the screening
of color charge by a scalar field  ``condensate".  Along a path in the $\b-\g$ plane
which continuously interpolates between the confinement-like and Higgs-like
regions, the two effects must smoothly morph into one another.\footnote{This fact may have
implications for the screening of adjoint representation (e.g.\ gluon) color charge in pure-gauge 
theories, since that effect is not essentially different from the screening of fundamental
representation color charge by a dynamical matter field.   Perhaps adjoint string-breaking by 
gluon pair-production can also be thought of as the screening effect of a gluon 
``condensate".}

     The dual abelian global gauge symmetry, probed by the monopole operator 
\rf{pisa-op} associated with dual-superconductivity,
has not yet been investigated in SU(2) gauge-Higgs theory.  However, there are 
already some indications, in G(2) gauge theory, that spontaneous breaking of the
 dual gauge symmetry is not necessarily accompanied by 
a change of physical state.  In G(2) lattice gauge theory there is known to be a point of rapid
crossover, where the plaquette action rises very sharply as $\b$ increases, but which does not
appear to be accompanied by an actual thermodynamic transition \cite{Pepe}. 
The monopole operator $\m$, or more precisely the logarithmic derivative $\rho=d\log(\m)/d\b$
of that operator, has been studied in G(2) gauge theory theory by Cossu et al.\ \cite{Cossu},
and preliminary numerical evidence suggests that the dual global gauge symmetry 
breaks at the crossover point, despite the absence of any actual change 
in the physical state at that coupling.  The signal of a transition in the monopole
operator, according to previous studies \cite{Adriano2}, is a large negative peak in $\rho$ at the
transition point, which grows with lattice volume, and this is found to be the case at the
crossover point at $\b=7/g^2=9.44$.  There is also a slight peak in $\r$  found at the 
deconfinement transition ($\b=9.765$ for $L_t=6$ lattice spacings in the time direction),
but this is tiny compared to the peak at the crossover point.  If there is indeed a transition
in $\m$ at the G(2) crossover coupling, that would be in line with what we have found for remnant 
gauge symmetries in Landau and Coulomb gauges: these symmetries break at points where 
there is no actual change of phase.
 
  There is still the question of whether there is any other symmetry 
which distinguishes confined from unconfined phases.  The answer hinges
on what is meant by the word ``confinement" (cf.\ ref.\ \cite{g2}).  If all it means is that the 
asymptotic particle states are color singlets, then there is really no
``unconfined" phase in gauge-Higgs theory, at any coupling, whose symmetry
could be contrasted with the confined phase.  If one chooses to define confinement
in this way, then the existence of a linear static quark potential is a separate, and
to some extent independent, issue.  There is, however an alternative 
definition of confinement, which we prefer: {\it Confinement is the phase of magnetic disorder.}
``Magnetic disorder" means the existence of vacuum fluctuations
strong enough to disorder, i.e.\ induce an area-law falloff in, Wilson loops at arbitrarily
large scales.  SU(2) gauge-Higgs theory is not in a magnetically 
disordered phase at any $\g>0$.   There
is always some cutoff length scale beyond which the large vacuum fluctuations, required
for the area-law falloff, are no longer found, and the vacuum state is magnetically ordered in the
infrared. (This is analogous to the concept of a massless phase: the phase does not exist
if the Euclidean propagator of the lightest particle state falls off exponentially at large distances, 
even if the falloff appears to 
follow a power-law up to some very large, but still finite scale.) A true magnetically-disordered vacuum
state, with magnetic disorder throughout the infrared region, is only found at $\g=0$, and there is
indeed a non-gauge symmetry which distinguishes the magnetically-disordered phase
at $\g=0$ from the ordered phase at $\g>0$.  This is the well-known global center symmetry.
The linear potential, linear Regge trajectories, and electric flux-tube formation are
only found, up to some finite distance scale, at small $\g$, where the center symmetry 
is only weakly broken (a situation labeled ``temporary confinement" in refs.\ \cite{GO1,g2}).   
As $\g\ra 0$ and center symmetry is restored, this finite scale goes off to infinity,
and magnetic disorder reigns throughout the infrared regime.   In theories where the center of
the gauge group is trivial, such as G(2) gauge-Higgs theory, a state of true magnetic disorder is never
reached, even at $\g=0$.

     Let us finally consider an SU(2) gauge-Higgs theory with the Higgs field in the adjoint 
representation.   In this case the
Lagrangian is invariant under center symmetry transformations, the symmetry is not broken
explicitly by the Higgs field, and this symmetry
can break spontaneously in certain regions of the coupling-constant space \cite{ucsc}.  The 
Fradkin-Shenker-Osterwalder-Seiler theorem does not apply in this case, and spontaneous center
symmetry breaking is a transition between two physically different phases, only one
of which is magnetically disordered.    The example is instructive.  Center symmetry
breaks spontaneously \emph{only} when there is a change in the physical state of the system, and
confinement $-$ understood as magnetic
disorder at all large scales $-$ is the phase of unbroken center symmetry. 
Global subgroups of a local gauge symmetry, on the other hand, can break 
spontaneously even when there appears to be no change of phase whatever, and their 
relevance to the confinement problem, in our opinion, remains to be firmly established.  
     
\acknowledgments{This research was supported in part by 
the U.S.\ Department of Energy under 
Grant No.\ DE-FG03-92ER40711 (J.G.) and by the
SEPAL/GK-12 Partnership Program under NSF
Grant No.\ DGE-0337949 (W.C).}

\end{document}